# Patterned growth of crystalline $Y_3Fe_5O_{12}$ nanostructures with engineered magnetic shape anisotropy

Na Zhu,[1] Houchen Chang,[2] Andrew Franson,[3] Tao Liu,[2] Xufeng Zhang,[1] E. Johnston-Halperin,[3] Mingzhong Wu,[2] and Hong X. Tang[1]

[1]*Department of Electrical Engineering, Yale University, New Haven, Connecticut 06511, USA*

[2]*Department of Physics, Colorado State University, Fort Collins, Colorado 80523, USA*

[3]*Department of Physics, The Ohio State University, Columbus, Ohio 43210-1117, USA*

Abstract

We demonstrate patterned growth of epitaxial yttrium iron garnet (YIG) thin films using lithographically defined templates on gadolinium gallium garnet (GGG) substrates. The fabricated YIG nanostructures yield the desired crystallographic orientation, excellent surface morphology, and narrow ferromagnetic resonance (FMR) linewidth (~ 4 Oe). Shape-induced magnetic anisotropy is clearly observed in a patterned array of nanobars engineered to exhibit the larger coercivity (40 Oe) compared with that of continuous films. Both hysteresis loop and angle-dependent FMR spectra measurements indicate that the easy axis aligns along the longitudinal direction of the nanobars, with an effective anisotropy field of 195 Oe. Our work overcomes difficulties in patterning YIG thin films and provides an effective means to control their magnetic properties and magnetic bias conditions.

Nanostructured ferromagnetic thin films have been considered as a promising platform in both longstanding fundamental studies of magnetic excitations and technological improvement of spintronic devices.[1–3] A large variety of magnetic nanostructures, such as nanoparticles, nanowires, and nanodots, have been successfully fabricated and widely studied for ferromagnetic metals such as permalloy and organic-based ferrites,[4–9] demonstrating the great potential of the utilization of these nanostructured devices in applications such as biological sensing,[10] data storage,[11,12] and logic devices.[13] Moreover, the precise dimension and morphology control offered by the nanostructured magnetic thin films would offer a promising protocol in engineering the unique properties of spin-wave excitations within the devices, paving the way for the advancement of future fundamental studies and practical applications of spintronic and magnonic devices.[14–17]

The shape anisotropy engineering of the ferromagnetic materials is attracting considerable interests for its applications in studying spin dynamics and building nanostructured microwave isolators and circulators.[18–20] However, in these works, the dimensions of the magnetic nanowires cannot be precisely defined and patterned, limiting the range of the operation frequency and the insertion loss of the devices. Therefore, precise shape anisotropy engineering via high-resolution geometric patterning is required to achieve fine tuning of the device magnetic properties, such as resonance frequency and coercivity.

Among all the magnonic media, ferrimagnetic insulator yttrium iron garnet ($Y_3Fe_5O_{12}$, YIG) attracts particular interests thanks to its extremely low damping.[21,22] High-quality single-crystal YIG films can be grown on lattice-matched substrates such as gadolinium gallium garnet ($Gd_3Ga_5O_{12}$, GGG),[23] by using liquid phase epitaxy (LPE), pulsed laser deposition (PLD), and magnetron sputtering.[21–26] However, the fine etching of single-crystal YIG thin films has long been a barrier for the study of magnonics and device applications of YIG at submicron dimensions. The previous methods for patterning microstructured single-crystal YIG films are wet chemical etching using phosphoric acid and ion milling, while using resists as masks.

Microstructured single crystal YIG has been fabricated via photoresist patterning and a following phosphoric acid wet etch to study the spin wave propagations in the magnonic crystal.[27] However, the wet etching process can only create relatively large structures and often leads to rough etching surfaces and etched steps that are not vertical with a significant slope.[28,29] Also, the YIG nanodisks have been used in spin wave studies by using resist patterning and ion milling etch.[30–32] But, the ion milling of YIG thin films, on the other hand, induces mechanical defects and the modifications of the magnetic properties of the films.[33,34] Another method for patterning single crystal YIG is selective-area growth, which is used for fabricating magnetooptic devices, offering limited resolution and device roughness.[35] Recently, anodic alumina oxide (AAO) membranes are used as the mask for patterning conical YIG nanoparticles on silicon substrates.[36] This interesting technique, however, is limited by the morphology of the AAO masks and cannot be extended to fabricate complex nanostructures such as nanoscale wires, rings, and disks.

In this paper, we demonstrate engineered magnetic shape anisotropy in YIG nanostructures formed by lithographical lift-off of YIG on lattice-matched GGG substrates followed by the high temperature annealing.[37–42] The patterned nanostructures retain very low magnetic loss with an peak-to-peak ferromagnetic resonance (FMR) linewidth as narrow as 4 Oe at 9.868 GHz. X-ray diffraction and surface metrology studies show that the formed nanostructures have desired crystallographic orientation and excellent morphology. The magnetic shape anisotropy is characterized by both hysteresis loop measurements and angle-dependent FMR measurements, and the data show that the patterned YIG film has been engineered to have a much larger coercivity compared with that of continuous films. The easy axis of the film is along the nanostrip's length direction with an effective anisotropy field of about 195 Oe. Our results demonstrate the great potential of utilizing patterned YIG thin films for both fundamental studies of nano-magnetism and the development of functional magnonic devices.

The device fabrication flow is schematically presented in Fig. 1(a). The process begins with a (111)-oriented GGG substrate. Bilayer PMMA resists (200-nm A3 and 1-μm EL 13) are first spun on a pre-cleaned GGG substrate, followed by sputtering deposition of a layer of 10-nm-thick gold to avoid electron charging effects on the insulating GGG substrate during the electron beam exposure process. The gold-coated sample is then exposed using an electron beam lithography tool (Vistec EBPG 5000+), and the gold layer is removed by gold etch after the exposure. The exposed sample is developed in a MIKB:IPA 1:1 solution to form a resist profile with deep undercuts resulting from the differential sensitivities of the two resists, which is critical for the final successful lift-off of YIG films after magnetron sputtering deposition.

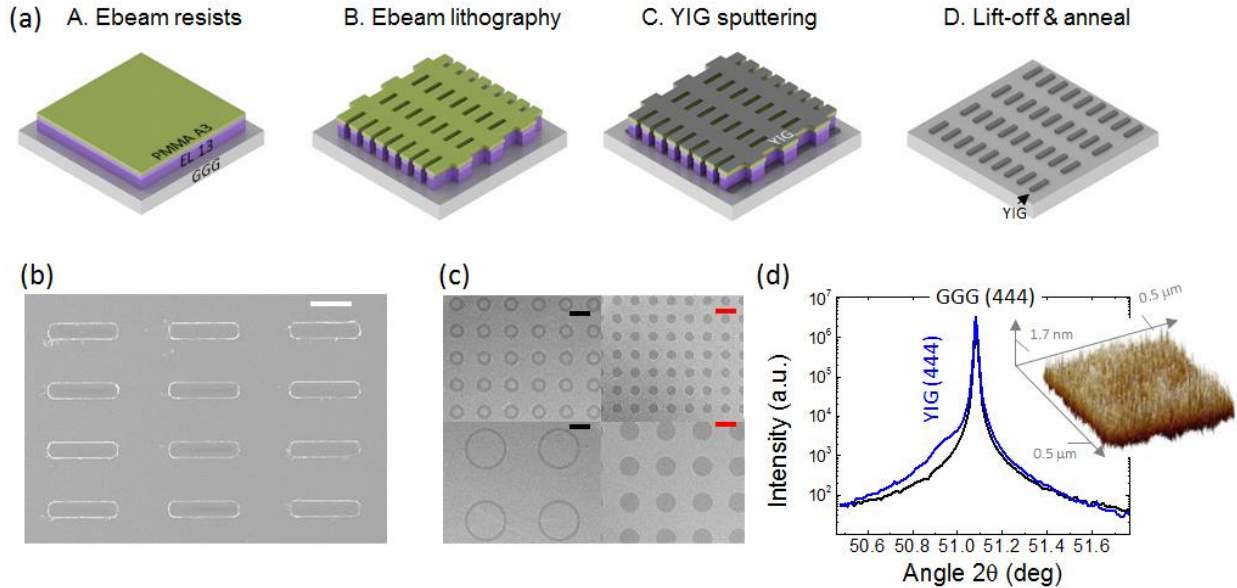

FIG. 1. (a) Schematic fabrication flow chart; (b) & (c) are the scanning electron microscope (SEM) images of the patterned YIG nanobars, rings, and disks. The white/black/red scale bar is 2/20/10 μm respectively; (d) X-ray diffraction spectra of the patterned

YIG nanobars and a pure GGG substrate. The inset is the atomic force microscope (AFM) surface image of the YIG nanobar sample.

A RF face-to-face magnetron sputtering chamber is used for our YIG film deposition. The deposition is carried out at room temperature with an Ar gas flow of 4 sccm, a gas pressure of 20 mTorr, and a sputtering power of 75 W. The thickness of the sputtered YIG film is around 75 nm, following by a lift-off process in acetone to remove the organic resist mask and the residual of sputtered amorphous YIG on the mask. Finally, the sample is annealed at 750 °C for 1 hour in a tube furnace with 10 Torr oxygen to form well crystallized structures. For the annealing process, the heating and cooling rates are about 10 °C/min and 2 °C/min, respectively.

The surface morphology of the fabricated nanostructured films is characterized by scanning electron microscopy (SEM), as shown in Figs. 1(b) and 1(c). To showcase our capability in patterning complex magnetic nanostructures, nanobars, rings, and disks with different dimensions are fabricated. All the patterned structures have clean boundaries and well-defined shapes with few defects. To confirm that the patterned YIG film is well crystallized, x-ray diffraction (XRD) measurements are performed on both the patterned YIG nanobars and the pure GGG substrate, and the spectra are shown in Fig. 1(d). The result reveals the existence of the YIG phase and no other phases, suggesting that the patterned YIG sample has a well crystallized structure with (111) orientation. Further morphological properties are analyzed by atomic force microscopy (AFM), as shown in the inset of Fig. 1(d), indicating the high-quality surface with a small root-mean-square surface roughness of about 0.47 nm. Note that the roughness value is an average over measurements on two different $0.5 \times 0.5$ μm areas, five times each. Our nanopatterned film via magnetron sputtering has a surface roughness similar to the surface roughness of (111) orientated GGG substrate and as-grown LPE YIG films (~0.4 nm).[43]

An array of nanobars is then used for the characterization of magnetic properties. The dimension of each individual YIG nanobar is 3 μm × 0.8 μm × 75 nm, and the total patterned area is around 2 mm × 2 mm with lattice spacing of 3 μm and 6 μm along the width and length directions, respectively. The nanobar has an approximate aspect ratio of 4:1 which leads to shape anisotropy induced by demagnetization along the transverse axis and the corresponding modification of the magnetic properties of the nanobar array as compared to continuous YIG thin films.

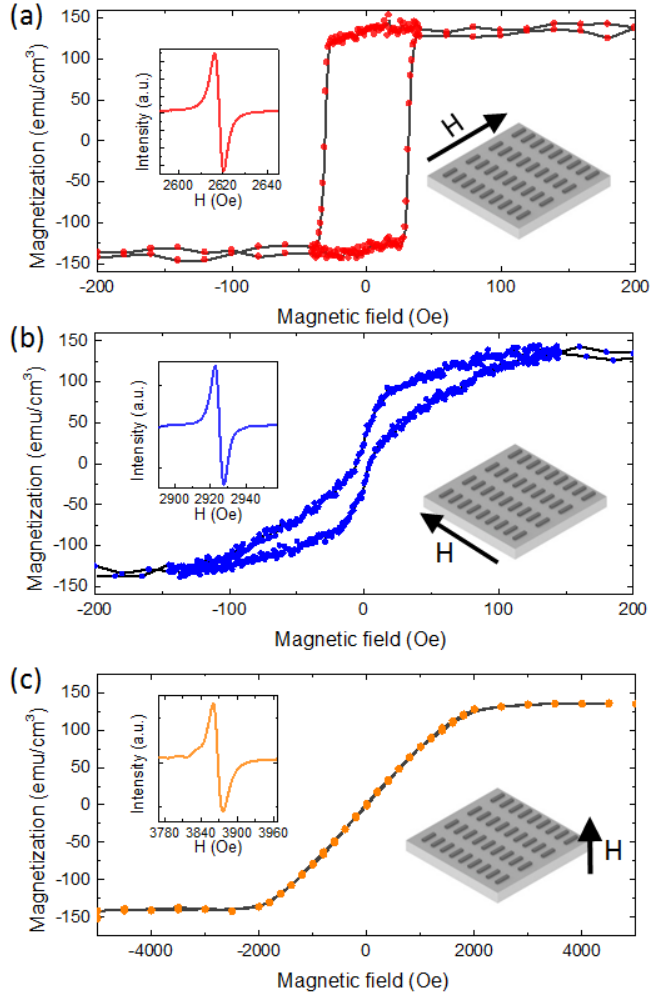

FIG. 2. (a)-(c) Room temperature hysteresis loops of the YIG nanobars measured at different magnetic field orientations. The insets are the FMR spectra measured at the corresponding field directions.

This can be seen in the hysteresis loops shown in Fig. 2, which are acquired at room temperature via vibrating sample magnetometer (VSM) with the external magnetic field varied between in-plane and out-of-plane directions. For the in-plane measurements, the external field is applied both parallel and perpendicular to the nanobar length direction. In contrast to the typical measurement results for continuous YIG thin films, the hysteresis loops for the in-plane geometry are significantly broadened and also show decreased saturation fields for fields applied parallel to the long axis of the nanobars. It is clear that the easy axis is aligned with the nanobar length direction, along which the magnetization saturates at a lower external field. Further, Fig. 2(a) shows that the coercive field is around 40 Oe when the field is parallel to the long axis. This coercivity is much higher than that for continuous films (~1 Oe).[37] This increase of the coercivity can be attributed to the increase of shape anisotropy due to geometry engineering of the nanobar structure.[20] When the external field is applied along the hard axis, the coercive field drops to 5 Oe, as shown in Fig. 2(b). Such an angular dependence can be qualitatively described by the Stoner-Wohlfarth model [44] in which a larger coercive field is expected when the external magnetic field is aligned along the easy axis of a hard magnet, while a smaller coercivity is expected when the external field is parallel to the hard axis.

As we can observe from the VSM measurements, the values of the saturation magnetization acquired from three VSM loops are slightly different. The measured saturation magnetizations when the external field is applied along longitudinal, transverse, and out-of-plane directions are $1736 \pm 70$ G, $1783 \pm 74$ G, and $1809 \pm 55$ G, respectively. The averaged value for the in-plane configuration is calculated to be 1759 G, which is close to the standard value of saturation magnetization of YIG (1760 G). The difference between the measured values and the standard value can be explained by the nonstoichiometry due to the existence of the chemical lift-off process in our fabrications.

The insets of Fig. 2 are the FMR spectra measured at room temperature by a Bruker electron paramagnetic resonance (ESR), which uses a microwave cavity with field modulation and lock-in detection techniques. The spectra show clean resonances with narrow linewidths, about 4 Oe for magnetic fields applied along the easy axis, indicating the high quality of the patterned film.

Compared with that of continuous films,[22,23,25] the relatively larger linewidth can be attributed to the magnon scattering from grain boundaries and void-like defects,[45] dimension distributions of the nanobars, residual materials around the boundaries, and inhomogeneous linewidth broadening.[22] The FMR linewidth can be further reduced by improving the patterning technique to offer smoother boundaries and more uniform dimension distribution. High order modes have been observed when the applied magnetic field is scanned over a larger range, showing an inhomogeneous magnetization distribution in the patterned structures.

To further study the magnetic shape anisotropy, we carried out angle-dependent FMR measurements at 9.868 GHz for an input microwave power of 0.22 mW. Figure 3(a) shows the spectra taken by stepping the angle of an in-plane magnetic field relative to the nanobar length axis from 0° to 360°. A clear shift of the main resonance is observed, confirming the magnetic anisotropy of the nanobar array. Another interesting feature of the spectra is the second resonance around 2767 Oe which has a much weaker angular dependence and corresponds to the FMR resonance of a reference square marker of 100 μm × 100 μm co-fabricated with the nanobar arrays. The relative resonance intensity difference between the nanobars and the square reference marker matches the total YIG volume ratio between them. In other words, this feature re-affirms that the shift of the FMR resonance field is dominated by the magnetic shape anisotropy field of the nanobars, not due to, for example, the magnetic crystalline anisotropy within the YIG film.

The angle dependence of the FMR resonances from the square marker can be used to extract the value of the magnetic crystalline anisotropy of the film. If we define $\theta$ and $\theta_H$ as the angles of the magnetization and the external field, respectively, with respect to the easy axis, the dispersion relation of the square marker for the in-plane magnetized configuration can be written as[46–49]

$$\frac{\omega}{\gamma} = \sqrt{(H\cos(\theta - \theta_H) + H_c \cos 4\theta) \times \left(H\cos(\theta - \theta_H) + H_\perp + H_c\left(1 - \frac{1}{2}\sin^2 2\theta\right)\right)} \quad (1)$$

, where the frequency $\omega$ equals $2\pi \times 9.868$ GHz, $\gamma$ is the absolute gyromagnetic ratio, $H$ is the external magnetic field, $H_\perp = 4\pi M_s(N_z - N_y)$, $H_c$ is the magnetocrystalline cubic anisotropy field, and $N_x$, $N_y$ and $N_z$ are the demagnetizing factors along $x$, $y$, and $z$ directions respectively. In our device configuration, where we assume $\theta \approx \theta_H$, the fitted value for cubic anisotropy field $H_c$ is $-5.4$ Oe. Using the first order cubic anisotropy constant for the single crystal YIG thin film $K_1 = -610$ J/m$^3$,[50] the magnitude of cubic anisotropy field is calculated as $|2K_1/M_s| = 87.5$ Oe. The difference between these two values indicates that large area marker has multiple magnetic domains, instead of a single one.

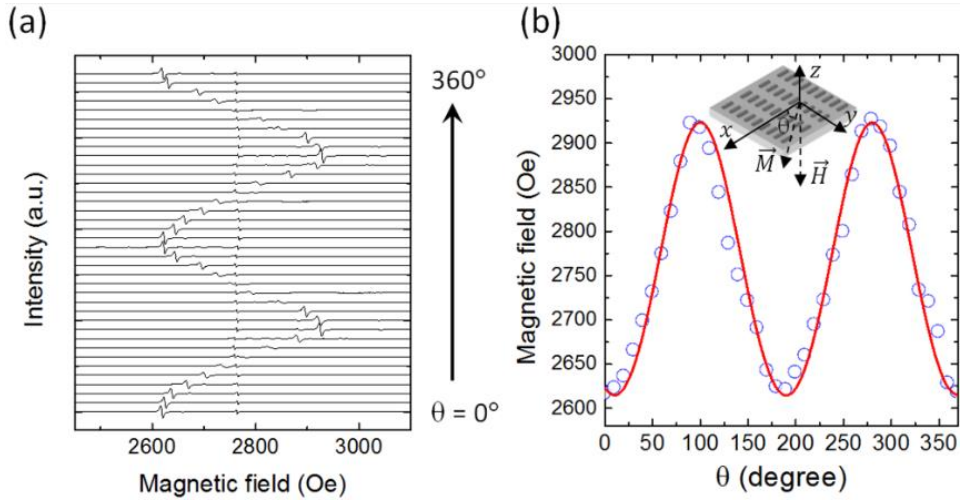

FIG. 3. (a) FMR spectra for patterned YIG nanobars when rotating an in-plane field; (b) FMR resonances field as a function of the in-plane field orientation is measured (symbol) and fitted via Eq. (1) in the main text (solid curve).

In order to extract the value of the effective magnetic anisotropy field, the center field of the FMR resonance is plotted as a function of the field angle in Fig. 3(b). The data exhibit a clear sinusoidal variation of the resonant field with regards to the angle of the applied field. Since the spacing between the patterned nanobars is as large as 3 µm, the magnetic dipole interaction and the exchange interaction between them can be neglected. As a result, we consider a magnetostatic model of individual rectangular nanobars and approximate them as a uniformly magnetized single-domain ellipsoid.[7,8,51–53] Considering the in-plane effective magnetic anisotropy field, the FMR frequency can be written as[46–49]

$$\frac{\omega}{\gamma} = \sqrt{\begin{array}{c}(H\cos(\theta - \theta_H) + H_A \cos 2\theta + H_c \cos 4\theta) \times \\ \left(H\cos(\theta - \theta_H) + H_\perp + H_A \cos^2\theta + H_c\left(1 - \frac{1}{2}\sin^2 2\theta\right)\right)\end{array}} \quad (2)$$

, where $H_A = 4\pi M_s(N_y - N_x)$ is the effective shape-induced in-plane anisotropy field. By substituting $H_c$ which equals −5.4 Oe, the values of $H_A$, $H_\perp$, and $\gamma$ are fitted by Eq. 2, and the fitted curve is plotted as the solid line in Fig. 3(b). The fitting yields $H_A$ = 195 Oe, $H_\perp$ = 1416 Oe, and $\gamma$ = 2.87 MHz/Oe. The fitted $\gamma$ value is close to the standard value which is 2.8 MHz/Oe.

The values of $H_A$ and $H_\perp$ can be calculated theoretically by using an effective demagnetizing tensor of the patterned nanobars. The nanobar with dimension 3 µm × 0.8 µm × 75 nm can be approximated as a general ellipsoid with demagnetizing factors along three principle axis (*x*,*y*, and *z*) to be 0.0096, 0.0881, and 0.91, respectively.[54] Using an average in-plane saturation magnetization of 1759 G we obtained in VSM measurements, the calculated $H_A = 4\pi M_s(N_y - N_x) = 138$ Oe is close to the fitted value. The difference can be explained by the coupling between nanobars, the nonstoichiometry of the deposited material, the dimension distribution of the nanobars and the simplification of the model by treating the nanobar as a general ellipsoid. Moreover, the calcuated $H_\perp = 4\pi M_s(N_z - N_y)$ is 1445 Oe, which is also very close to the fitted value.

In conclusion, we have demonstrated the synthesis of high-quality YIG nanostructures on lattice-matched GGG substrates via room temperature magnetron sputtering and lithography. Structures with different geometries and dimensions have been realized. In particular, the structural and magnetic properties of YIG nanobar devices are systematically studied. The devices are demonstrated to have crystalline structures and narrow FMR linewidths. Hysteresis measurements and FMR characterizations reveal that geometry engineering of the YIG nanobars controls magnetic properties such as shape anisotropy and coercivity, allowing significant improvement as compared to the behavior of un-patterned YIG films. Our results establish the feasibility of precisely tuning the magnetic properties of high-quality YIG thin films and present more opportunities for the utilization of structured YIG for spintronic and magnonic device applications, such as spin torque transfer, spin dynamics in coupled nanopatterned YIG devices, and the stabilization of the magnetic order via shape anisotropy at low temperature.

We acknowledge funding support from an LPS/ARO grant (W911NF-14-1-0563), Air Force Office of Scientific Research (AFOSR) Multidisciplinary University Research Initiative (MURI) grant (FA9550-15-1-0029), Defense Advanced Research Projects Agency (DARPA)

Mesodynamics Systems (MESO) program, the National Science Foundation (DMR-1507775), and the Packard Foundation. Facilities used were supported by the Yale Institute for Nanoscience and Quantum Engineering (YINQE), the Yale cleanroom, and the NanoSystems Laboratory (NSL) at The Ohio State University. The work at CSU was supported by SHINES, an Energy Frontier Research Center funded by the U.S. Department of Energy (SC0012670); the U.S. National Science Foundation (EFMA-1641989); the C-SPIN, one of the SRC STARnet Centers sponsored by MARCO and DARPA; and the U. S. Army Research Office (W911NF-14-1-0501). We would like to thank Michael Power and Christopher Tillinghast for the assistance in device fabrication.